# No Code AI: Automatic generation of Function Block Diagrams from documentation and associated heuristic for context-aware ML algorithm training

Oluwatosin Ogundare, Gustavo Quiros Araya, Yassine Qamsane
Siemens Technology
Princeton, NJ, USA
gustavo.quiros@siemens.com

*Abstract*— **Industrial process engineering and PLC program development have traditionally favored Function Block Diagram (FBD) programming over classical imperative style programming like the object oriented and functional programming paradigms. The increasing momentum in the adoption and trial of ideas now classified as "No Code" or "Low Code" alongside the mainstream success of statistical learning theory or the so-called machine learning is redefining the way in which we structure programs for the digital machine to execute. A principal focus of "No Code" is deriving executable programs directly from a set of requirement documents or any other documentation that defines consumer or customer expectation. We present a method for generating Function Block Diagram (FBD) programs as either the intermediate or final artifact that can be executed by a target system from a set of requirement documents using a constrained selection algorithm that draws from the top line of an associated recommender system. The results presented demonstrate that this type of No-code generative model is a viable option for industrial process design.**

## I. Introduction

In industrial control, Function Block Diagrams (FBD) are widely used for designing control systems software implemented in Programmable Logic Controllers (PLC). FBD programming is one of the standard PLC programming languages defined in the IEC 61131-3 standard [1] and its extension, the IEC 61499 standard [2]. Currently, there is a fast-growing number of methods and heuristics for code generation from requirements documentation. Many of these solutions rely on the simplicity of FBDs to support the idea that code generation from documentation is industrially viable. Even in cases where scaffolded object oriented, or functional code are generated as the final artifact, FBDs are sometimes generated as the intermediate or secondary artifact.

Code generation in current industrial control practice is typically based on either heuristic methods or formal methods. Heuristic methods intuitively translate the requirements from documentation into control code based on the practitioner's ingenuity [3][4]. For small size systems, this direct interpretation of the requirements into control code may yield practical solutions, however, as the system size becomes larger, the method becomes insufficient and error prone. To assist the control practitioners in their task, formal methods can be exploited to automatically generate control code.

A specification language for control code generation based on linear temporal logic is widely adopted in the industry [5]. Event-B formalism is also widely used to design PLC control code and requires formal proofs that the designed code satisfies completeness, consistency, precision, and correctness throughout the modeling and refinement process [6]. Other methods combine several formal languages with PLC standard languages, such as automata and GRAFCET/SFC [7], Petri nets and Ladder diagrams [8][9], and Max Plus Algebra with Petri Nets [10][11].

Formal methods outperform conventional heuristic methods as they use well designed algorithms to automatically generate control code with less effort and time. However, from a practical standpoint, formal methods require the control practitioner to have strong knowledge of advanced mathematical concepts used to either model the requirements from documents or the system/process to control or both. Most control practitioners in industry do not have the necessary knowledge to work with formal methods. Thus, there is a need for additional methods for translating the requirements to mathematical specifications used by formal methods algorithms to generate control code. Recently, Machine Learning (ML) approaches has been applied to simply the modelling interface and minimize design complexity. For example, a tool and technique based on deep learning for translating informal requirements into formal Signal Temporal Logic (STL) was shown to perform as well as formal methods [12]. Furthermore, active learning has been used successfully to synthesize controllers for unknown plant models [13]. Active automata learning has been introduced in [14] to automatically learn formal automata models. It has found many applications, ranging from security analysis, testing, verification, and synthesis. The combination of formal algorithmic methods with learning algorithms can infer more expressive models with less effort from the practitioner. This paper explores the generation of executable FBD programs from documentation under the No Code engineering paradigm in industrial and process-centric domains by first generating a constrained solution space from the reference documentation and selecting the function blocks that maximizes the conditional probabilities computed from the same utility function that may serve independently as a recommender system.

## II. AXIOMATIC DEFINITION OF THE CONTEXT-AWARE INDUSTRIAL PROCESS DESIGN USING FBD PROGRAMMING

**Axiom 1.** $\forall g \in G$ there is an associated utility function, $U$, that characterizes the industrial function of $g$, and a function $\pi$, that assigns a series of sets $\{S_j\}$ such that $S_j$ is composed of design elements in $G$ but allows repetitions, such that the following holds

$$\exists g_1 \in G : (U_{g_1}(g_1) \land \forall t_1 \in T_j : U_{g_1}(t_1) \rightarrow g_1 = t_1)$$

Theoretically, $j \rightarrow \infty$, since symbols can repeat arbitrarily, but there is a practical computational limit to the structure of FBD programs. Informally, $\pi$ can be viewed as the function that assigns the symbols needed to create an FBD program from $G$ and sometimes a symbol is used more than once. The total number of function blocks or more generally symbols a FBD program contains without reference to uniqueness is described by $S_j$. From the viewpoint of process (design) engineers, having been assigned $S_j$, they define the arrangement of the symbols or function blocks in $S_j$. A visual analog of $S_j$, is to think of it as a set of puzzle pieces that is meant to be assembled. Using this analogy, $G$ would be unique pieces in $S_j$. In the domain of factory automation, $\pi$ is almost always defined by the requirements documentation, for example, the number of Boolean gates/blocks might correlate to the number of valves in a Process & Instrumentation Diagram (P&ID).
The FBD design function, informally known as the puzzle arrangement function or more formally as the action of a design engineer over a set of FBD symbols/blocks is modelled by the function $\phi$, and defined subsequently.

**Axiom 2.** $\phi_i : S_j \rightarrow \mathbf{f}_i$ such that $\mathbf{f}_i$ is the set of all $i$ permutations of $S_j$. $\mathbf{F}$ is the set of all known permutation of $S_j$.

$$\mathbf{f}_i = \{f_{i,k}\}$$
Where $k = 1, \ldots, |S_j|!$
$i = \{1, \ldots, |S_j|\}$

$T_j$ is normally distributed i.e., $T_j \sim N(\mu, \sigma^2)$ as a practical consideration because the design elements in the FBD toolbox are not equally likely. Some function blocks are used more than others and as such are more likely to occur in an instance of an FBD program.
$\psi_i$ is defined over $\mathbf{f}_i$ as the probability distribution of the set of all $i$ permutations of T.
$$\psi_i : f_{i,k} \rightarrow (0,1)$$

$f_{i,k}$ can be informally referred to as an instance of an FBD program or a set of symbols. For example, the simplest FBD program is in $\mathbf{f}_1$ (set of all FBD programs with a single symbol). Similarly, the next level of complication admits FBD programs with 2 symbols/function blocks which is $\mathbf{f}_2$. Along these lines, $f_{2,1}$ is a valid 2-symbol FBD program in $\mathbf{f}_2$ and $f_{2,2}$ would refer to another valid 2-symbol FBD programs or set of symbols/function blocks in $\mathbf{f}_2$.
$\psi_i$ is constrained by the axioms of probability and that the $\mathbf{f}_i's$ are exhaustive, i.e., if the last choice is in $\mathbf{f}_{i-1}$ the next choice is guaranteed to be $\mathbf{f}_i$ and the next in $\mathbf{f}_{i+1}$.
Simply put, given that $\mathbf{f}_{i-1}$ is realized

$$\Pr(\mathbf{f}_i) = \begin{cases} 1, & \text{If the engineer makes another selection} \\ 0 & \text{otherwise} \end{cases}$$

However more specifically, we are only interested in the probability distribution of all valid $i$-symbol FBD programs given a particular $i-1$ symbol has been realized. We estimate the conditional probability as follows

$$\Pr(\mathbf{f}_i = f_{i,k} \mid \mathbf{f}_{i-1} = f_{i-1,b})$$
$$= \frac{\boldsymbol{Pr}(\mathbf{f}_{i-1} = f_{i,b} | \mathbf{f}_i = f_{i,k}) \psi_i(f_{i,k})}{\psi_{i-1}(f_{i-1,b})}$$

Consequently, the recommended next step is the one that maximizes the conditional probability.

$$\max_k \{ \Pr(\mathbf{f}_i = f_{i,k} \mid \mathbf{f}_{i-1} = f_{i-1,b}) \}$$

Where $k, b = 1, \ldots, i!$
$i > 1$

The initial recommendation follows the distribution of $T_j$ and subsequently the distribution of the conditional probability is used, so $i > 1$

Intuitively, $\mathbf{f}_i - \{f_{i,k}\}$ represents the set of all alternate configurations with $i$ design elements chosen from $T_j$ with a preset $i - 1$ configuration. These alternate configurations follow a conditional probability distribution and can be useful for a multi-option recommender system with the alternates ranked by the conditional probability.
Since generating all permutation of a set of function block symbols is the factorial of the total number of FBD programming symbols and generally known to be computationally NP-hard, that is, the computational complexity grows exponentially with the size of the set, in this case, the length of the FBD program symbol set especially since there is no differentiation between the connectors and the function blocks themselves. In graph modelling terms, our approach abstracts away differentiation between edges and vertices. All elements used in building the FBD program are simply referred to as function blocks or more generally symbols. We conceive a heuristic that allows the assignment of ZERO probability to large chunks of the permutation set which consequently changes the context learning probability distribution from uniform over the set of all possibilities to a distribution to a distribution that is geometrically stochastic.

## III. FIONA HEURISTIC FUNCTION

The forward Permutations are bijections but in industrial FBD program design, not all transpositions of the symbols/function blocks are structurally valid. Subsequently, we introduce the **Functionally Organized Normalized Arrangements (FIONA)** permutation function as a heuristic function for generating context defining datasets.

Consider a set of two symbols, all the permutations can be generated a single transposition function, $\sigma$. However, for a set of 3 symbols, we need 3 transposition functions, $\sigma_1, \sigma_2, \sigma_3$ to generate all permutation of the set. We denote $\sigma_1$ as the transposition that fixes symbol 1 and exchanges symbol 2 & 3. Similarly, $\sigma_2$ fixes Symbol 2 and exchanges the others and $\sigma_3$ fixes Symbol 3 and exchanges the others. Rigorously, the FIONA heuristic function, $\sigma_{fiona}$ is a subgroup of the permutation group of n-symbols equipped with an image and a kernel.

At every iteration $i \in \{1, 2, 3, \ldots, k\}$ we define a set of exclusions $E = \{e_1, e_2, e_3, \ldots, e_k\}$

We select the set $S_i = \{s_{i,j} \in \ ^np_2 \mid s_{i,j} \cap e_i = \emptyset \}$

We assume that $S_i$ has a Cumulative Distribution Function (CDF) that is uniform on $(0, 1)$. We derive a probability density function from the historical distribution of the function blocks in various FBD programs. Intuitively, some function blocks will be more likely to appear in more FBD programs, such that for every $s_{i,j}$, there is an associated conditional probability, $\Pr(s_{i,j})$. Such that a symbol is paired with its probability as follows ($s_{i,j}, \Pr(s_{i,j})$)

## IV. FIONA LONG SEQUENCE BUILDING OF CONTEXT TRAINING DATA

We introduce $\gamma_i$ such that it randomly selects a value for the Cumulative Distribution Function (CDF) at iteration $i$, to the order of $S_i$, such that

$$\gamma_i = \int_0^1 f(k)$$

$f(k)$ is the density function evaluating the likelihood of the symbol with k-index(value)
$j, k = \{1, \ldots n \ (total \ number \ of \ symbols)\}$

This will result in algebra with parameter, $j$, as the unknown and the solution will follow from first principles.

The sequence generating function, $\tau_{fiona}$, such that at every iteration, i
$\tau_{fiona}: (i, j) \rightarrow \{s_{i-1,k} \cup s_{i,j}\}$

## V. RESULTS ON FIONA TRAINING HEURISTIC

Informally, every time $\pi$ makes a selection in a permutation set, $s_i$, we create a subproblem of what is possible in $s_{i+1}$

Formally, we learn the image of $\pi_j(s_i)$ in $s_{i+1}$. $\pi_j$ is the specific selection strategy that produces a unique FBD program $(FBD_j)$. We also learn what actually happened in $s_{i+1}$ which is $\pi_j(s_{i+1})$, the methodology assigns higher weights to $\pi_j(s_{i+1})$, which differentiates this type of learning from the other possibilities in $s_{i+1}$ based on the previous selection $\pi_j(s_i)$

To simplify the learning problem, we assign a different instance of the chosen ML model, in this case, a Recurrent Neural Network (RNN) the task of predicting $\pi_j(s_i)$, i.e., the action of a selection strategy $\pi_j$ in $s_i$. How much of the prior selections and images accumulated in each RNN via LSTM cells as well as computational complexity constitute a practical design problem.

The following figures shows the error surface when training across a large collection of FBD programs segregated by design steps, i.e., we select a set of the first 2 function blocks across n-FBD programs, then we select the first 3 function blocks across the same n-FBD programs, we select the first 4 functions across the same n-FBD programs and so on. The main idea is to see how well an ML model performs in minimizing the error on the training set having been trained on more generic context data.

We begin by training an RNN with 50 LSTM cells on learning the relationship between 1 symbol to 2 symbols transitions, then subsequently we train a different instance of the RNN with the same configuration on 2 symbols to 3 symbols transitions and so on. Finally, we perform the same training with the same dataset on RNNs with the same configuration but pre-trained on context data focused on identical transitions and visualize the error surface in Fig 1 – 6.

1. Training 1 Symbol to 2 Symbols Transition

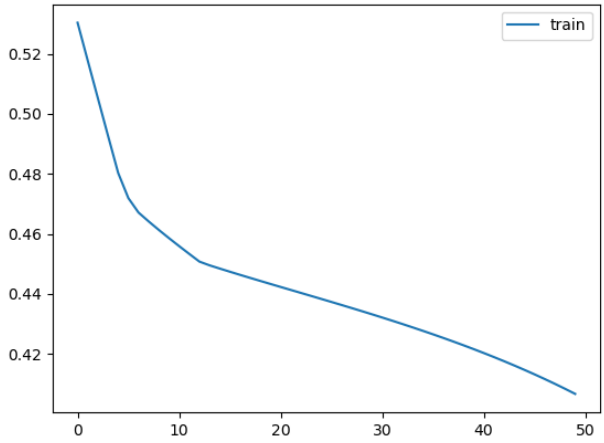

Figure 1. Plot showing the error surface of 1-2 transitions after 50 epochs.

2. Training 2 Symbols to 3 Symbols Transition

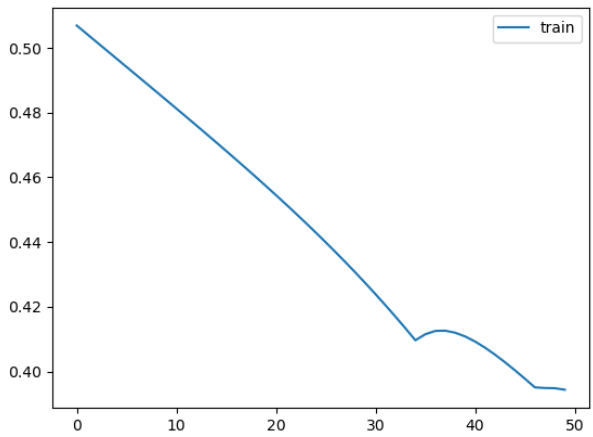

Figure 2. Plot showing the error surface of 2-3 transitions after 50 epochs.

3. Training 3 Symbols to 4 Symbols Transition

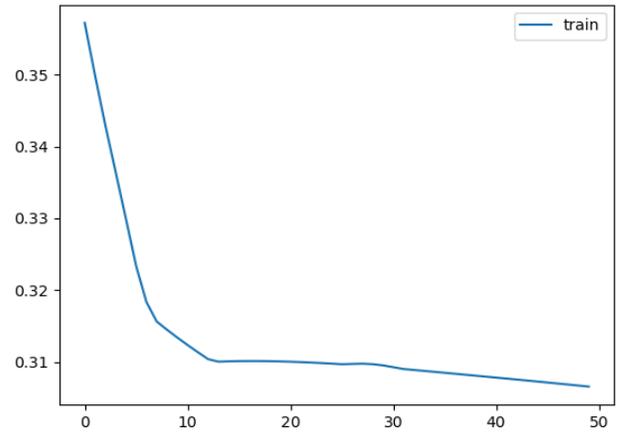

Figure 3. Plot showing the error surface of 3-4 transitions after 50 epochs.

In Figure 4-6 below, we visualize the error surface over the same dataset using RNN pre-trained on context data.

1. Pre-trained RNN Training 1 Symbol to 2 Symbols Transition

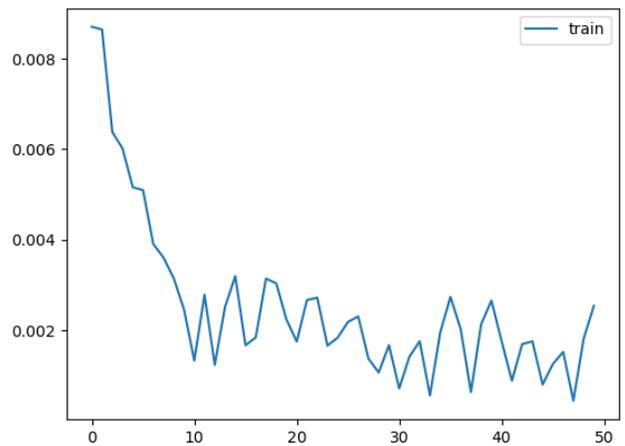

Figure 4. Plot showing the error surface of 1-2 transitions after 50 epochs.

2. Pre-trained RNN Training 2 Symbols to 3 Symbols Transition

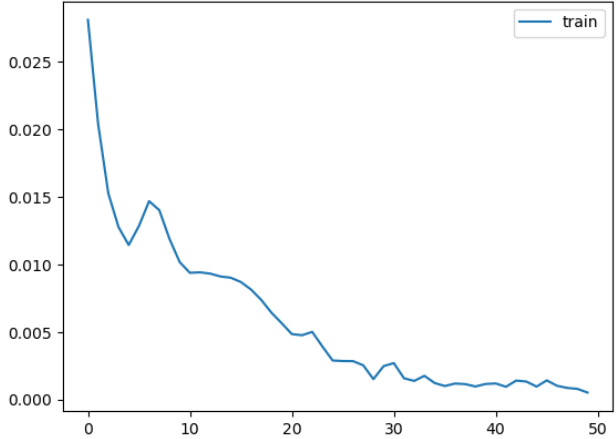

Figure 5. Plot showing the error surface of 2-3 transitions after 50 epochs.

3. Pre-trained RNN Training 3 Symbols to 4 Symbols Transition

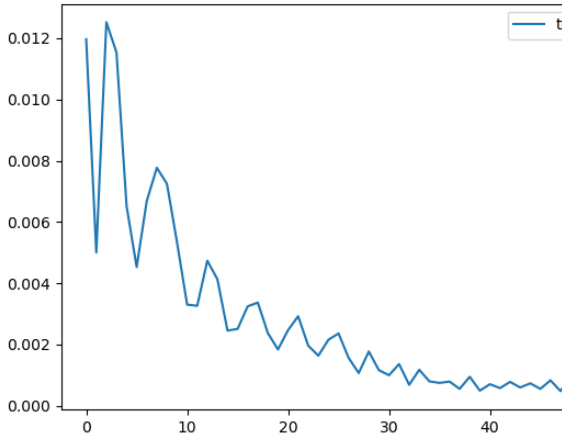

Figure 6. Plot showing the error surface of 3-4 transitions after 50 epochs.

## VI. AUTOGENERATION OF FBD PROGRAMS

We introduce the notion of the action model as a time synchronized model that produces a result at a discrete design time step. Within the context of a FBD program, an action model at t=0, performs the initial selection of the first symbol in the FBD program design. Another action model selects the second function block or more generally, the next symbol at the next design time step. The design time is a discrete random variable that only takes a value when a design action is performed, i.e., when a symbol or function block is selected and added to the FBD program.

In this paradigm, a FBD program is viewed as the realizations of a collection of time synchronized action models. An action model, A at time t, learns the probability distribution of the universe of symbols and selects the symbol or function blocks that maximizes the conditional probability.

Suppose for every set of symbols $U = \{u_i\}$, we define the notion of an FBD program, $S$, such that at every design step, an action model

$$A_t: U \rightarrow S^k$$

Where t = 1,...,N and are discrete design time steps

K indexes the set of possible FBD programs over the U

We claim that the variation observed in the different plant and FBD programs with identical goals are a function of engineering preferences, experiences, and design guideline interpretations and that this variation is encapsulated in the statistical distribution of $S^k$ over the $U$ for a specific task.

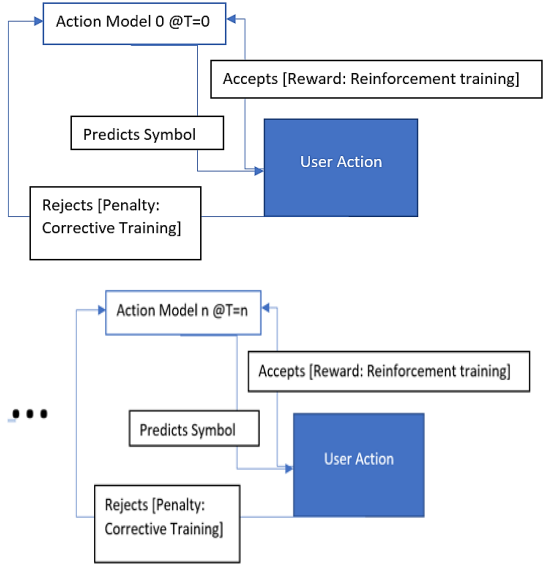

Figure 7. Training of an action model in FIONA-4 for singular action at unique time steps

We introduce the FIONA-4 as shown in Figure 7 Machine Learning (ML) system that localizes Neural Network models (RNN), and trains them on the action of the engineers working on similar problems to learn this variation. Every cluster of localized Neural Network models assigns a training task at each design time step, $t$, to a unique Neural network model, whose job will be to learn the probability distribution of $U$ over $S_{t-1}^k$ so that it can later perform the function of selecting the best $u \in U$ at time $t$, for an arbitrary "out of sample" task given that the state of the design at time $t-1$ is $S_{t-1}^{arb.\ task}$.

The goal the Action model is defined more clearly as follows:

$$A_t: argmax\ (U \mid S_{t-1}^k) \rightarrow S_t^k$$

Refer to figure 7. Above for a schematic of each time synchronized action model training.

FIONA-4 ML federates the localized cluster of Neural Network models and uses these as a basis for automatic FBD program generation system by auto-accepting the selection of the action model $A_t$.

## VII. CONCLUSION

A novel method for training heuristic was shown and its overall impacted on the convergence of training on customer data is presented. As well as using the trained RNN as a generative model recursively to create FBD programs using the predictions from the RNNs or the more generally the function block that maximizes the conditional probability based on the state of the design or the prior model selections.